\documentstyle[prd,aps]{revtex}
\newcommand{\beq}{\begin{equation}}
\newcommand{\beqn}{\begin{eqnarray}}
\newcommand{\eeq}{\end{equation}}
\newcommand{\eeqn}{\end{eqnarray}}
\begin{document}
\draft
\preprint{HUTP-97/A041; hep-ph/9707516}
\title{Resonance Structure for Preheating with Massless Fields}
\author{David I. Kaiser}
\address{Lyman Laboratory of Physics, Harvard University, Cambridge, MA
02138 USA}
\date{July 1997, revised November 1997}
\maketitle
\begin{abstract}
We extend recent work on the resonance structure for post-inflation
reheating, providing an analytic treatment for models in which both the
inflaton and the fields into which it decays are massless.  Solutions are
derived which are valid for either a spatially-flat or spatially-open
metric.  Closed-form solutions are given for the characteristic exponent,
which measures the rate of particle production during preheating.  It is
demonstrated that in certain ranges of parameter space, the maximum values
of the characteristic exponent in an open universe are several times
greater than the maximum values in a spatially-flat universe.  It is
further demonstrated that the solutions found here by means of a simple
algebraic construction match the two previously-known exact solutions,
which were derived in terms of special functions.
\end{abstract}
\pacs{PACS 98.80Cq, 04.62,+v \hspace*{3.5cm} Preprint HUTP-97/A041,
hep-ph/9707516}
\vskip2pc

\section{Introduction}
\indent  In this paper we extend recent work by Greene, Kofman, Linde, and 
Starobinsky on the resonance structure for models of post-inflation
reheating in which
both the decaying inflaton and the fields coupled to it are
massless. \cite{GKLS}  Their work treats the case of massless fields in a
spatially-flat background spacetime.  Yet there is increasing evidence to
suggest that we live in an open universe, and the recent models of open
inflation present an interesting way to reconcile an open universe with an
early phase of cosmological inflation. \cite{openI}  The formalism of
\cite{GKLS} cannot be used directly to study the preheating dynamics
following a phase of open inflation.  Here we extend the work of
\cite{GKLS} to treat preheating in an open universe, and find both
qualitative and quantitative changes in the spectra of produced quanta
between spatially-open and spatially-flat models.  In particular, when the
initial value of the oscillating inflaton field is of the same order of
magnitude as the associated curvature scale, there will be {\it no}
resonant amplification of any modes -- a result with no correlate in the
corresponding spectra for preheating in a spatially-flat universe.
Furthermore, characteristic exponents can reach larger values in an open
universe than in a spatially-flat one, signalling a quicker rate of
resonant particle production.  Because of the conformal invariance of the
massless theory, the extension
provided here also allows study of preheating in Minkowski spacetime with
massive fields, with or without explicit symmetry-breaking terms in the
potential.  With these extensions, the elegant and powerful methods
introduced in \cite{GKLS} may be applied to a much wider class of
interesting models. \\
\indent  The case of massless fields in an expanding spacetime admits
a self-consistent 
analytic treatment for all three of the time-dependent quantities
involved:  the oscillating inflaton, the resonantly-amplified
decay-product field, and the scale-factor for the background spacetime.
Whereas previous authors have found closed-form solutions for such
preheating systems in terms of special functions for two particular ratios
of the relevant couplings
\cite{DBan,DK97}, the authors of \cite{GKLS}
produced a method
for finding solutions for a far broader class of couplings by means of a
much
more simple, algebraic construction.  In addition to extending their work
to treat preheating in an open universe, we also demonstrate here that
the Floquet index, which measures the rate of resonant
production of decay-product quanta, can often be written in closed form,
rather than in terms of the \lq\lq auxilliary functions" found in
\cite{GKLS}, which are defined as definite integrals and evaluated
numerically.  Finally, we demonstrate 
explicitly that the solutions found by the algebraic construction of
\cite{GKLS}, with only minor modifications, match the exact results found
previously in terms of special functions. \\
\indent  As is by now well known, linear second-order differential
equations with
purely periodic coefficients have solutions which obey Floquet's theorem:
the solutions behave as $X (t + 2\omega) = X (t) \exp (i F t)$, where
$2\omega$ is the period of the coefficients in the governing equation of
motion, and $F$ is the Floquet index (also known as the characteristic
exponent).  While the form of $F$ will depend on model parameters, it will
be independent of $t$.  These solutions can always be rewritten in the
form $X(t) = P(t) \exp (\mu_\kappa t)$, where $P(t)$ is a periodic
function
with period $2 \omega$, and $\mu_\kappa$ is simply related to $F$ (see,
{\it e.g.}, \cite{Ince}).  Obviously, when $\mu_\kappa$ has non-zero real
parts, the solutions develop exponential instabilities.  Such solutions
are said to lie within \lq\lq resonance bands" or \lq\lq instability
regions," the dependence of which on
model parameters can often be written explicitly based on an analysis of
the form of $\mu_\kappa$.  Floquet's theorem lies at the heart of the
study of preheating with massless fields, since, for early times after the
inflaton has begun oscillating around the minimum of its potential, the
equations of motion for any fields coupled to it assume the form of
second-order differential equations with periodic coefficients.  It is
therefore imperative to develop efficient and accurate methods for
evaluating $\mu_\kappa$ for a broad class of interesting models.  This is
the main goal of this paper. \\
\indent  In Section II, we present the models to be studied and their
dynamics.  We also discuss the question of appropriate vacuum
states with respect to which the preheating production of quanta  
should be measured.  Section III presents the extension of the method of 
\cite{GKLS} for the new cases, and includes explicit solutions of the  
equations of motion valid for early times (before back-reaction becomes
significant).  In Section IV we demonstrate how to re-write the
$\mu_\kappa$-parameters in closed form for many cases, which facilitates
both their
numerical evaluation and their comparison with known solutions. 
Appendix A includes an explicit demonstration of the equivalence of the
solutions $X(t)$ found in Section III with solutions found by other means.
In Appendix B it is shown that the $\mu_\kappa$-parameters calculated in
Section IV likewise match the exact solutions found previously.

\section{Dynamics of the Model}
\indent  The effective Lagrangian density we will study involves two    
fields:  an inflaton $\phi$ and a decay-product field $\chi$, both of
which are massless and have minimal couplings to the Ricci curvature
scalar.  For the field content of the model, we may then write:
%%%%%%%%%%%%%%%%%
\beq
{\cal L} = - \sqrt{-g} \left[ \frac{1}{2} \left(
\partial_\mu \phi \right)^2
+ \frac{1}{2} \left( \partial_\mu \chi \right)^2 + \frac{\lambda}{4}   
\phi^4 + \frac{g^2}{2} \phi^2 \chi^2 \right]  .
\label{Lag}
\eeq
The spacetime near the beginning of the preheating epoch will assume the
form of an FRW metric; for the usual models of inflation, this will be a 
spatially-flat metric (with $K = 0$), while for the newer models of open  
inflation, this will be a spatially-open metric (with $K = -1$):
%%%%%%%%%%%%%%%%%%
\beqn
\nonumber  ds^2 &=& -dt^2 + a^2 (t) \> h_{ij} dx^i dx^j , \\
\nonumber h_{ij} dx^i dx^j &=& d {\bf x}^2 \>\>\> (K = 0) , \\
&=& dr^2 + \sinh^2 r \left( d\theta^2 + \sin^2 \theta d \phi^2 \right)
\>\>\> (K = - 1) .
\eeqn
For either case, the equations of motion for the fields become
%%%%%%%%%%%%%%%%%%%%%
\beqn
\nonumber  \ddot{\phi} + 3 H \dot{\phi} + \lambda \phi^3 + g^2 \chi^2 \phi
&=& 0 \\
\ddot{\chi} + 3H \dot{\chi} - \frac{1}{a^2 (t)} {\bf L}^2 \chi + g^2
\phi^2 \chi &=& 0 ,
\eeqn
where dots denote $\partial / \partial t$, $H \equiv \dot{a}/a$, and ${\bf
L}^2$ is the comoving spatial Laplacian operator.  We have assumed that
the slow-roll field $\phi$
is spatially homogenous; that is, we will ignore the spectrum of inflaton
fluctuations for now.  As
demonstrated in \cite{GKLS}, the preheating production of
inflaton quanta from a massless inflaton with $\lambda \phi^4$ coupling is
anomalously inefficient, as compared with the preheating production of a
distinct (massless) boson field.  The spectrum of inflaton fluctuations
may also be studied in this formalism by associating $\chi$ with
$\delta \phi$, and assigning $g^2 = \lambda$ in the 
large $N$ limit of the $O(N)$ approximation, or $g^2 = 3 \lambda$ in the
Hartree approximation.  (See, {\it e.g.}, \cite{GKLS,DBan}.) \\
\indent  We may now promote the field $\chi$ to a Heisenberg operator.
Details of this expansion, and of the canonical quantization of the
resulting operators for both the cases $K = 0$ and $K = -1$, may be found 
in \cite{DK97}.  For the $K = -1$ case, we will only track subcurvature
modes here; the possibility of amplifying supercurvature modes during
preheating in an open universe will be treated elsewhere. \cite{DKinprep}
The quantum $\chi$ field may then be expressed as a sum over modes and
associated creation and annihilation operators, with
%%%%%%%%%%%%%%%%
\beq
\hat{\chi} (t,\> r,\> \Omega) = \int d \tilde{\mu} \left[
\chi_{p \ell m} (t, \> r, \> \Omega)\> \hat{a}_{p \ell m} +
{\rm H.c.} \right] .
\eeq
Here $d \tilde{\mu}$ is the appropriate measure for the integral,
including relevant sums over $\ell$ and $m$ (the specific form 
depends on whether $K = 0$ or $K = -1$, see \cite{DK97}),
$\hat{a}_{p \ell m}$ is a canonical time-independent annihilation
operator, and \lq\lq H.c." denotes a Hermitian conjugate. \\
\indent  In order to study the non-linear dynamics, we will make use
of a
Hartree factorization.  Because we will be studying the evolution of the  
fields before any back-reaction becomes significant, the details of the
resonance structures for the model of equation (\ref{Lag}) will in fact
remain independent of the specific 
approximation scheme invoked to study the 
non-linearities (though details concerning how long the
preheating phase will last do depend on the choice of such schemes).  In
our case, the Hartree approximation entails replacing the $g^2 \chi^2$
term in the equation of motion for $\phi$ with its vacuum expectation
value, $g^2 \langle \hat{\chi}^2 \rangle$.  This quantity gives a measure
of the growth of the back-reaction upon the oscillating $\phi$ field due
to the resonant shift of energy into $\chi$ modes.  Because of spatial
translation invariance, this quantity can only depend upon time.  We will
hence study the coupled system for early times, when this back-reaction
term may be neglected relative to the tree-level terms. \\
\indent  The equations of motion become much more simple if we rewrite
them
in terms of conformal time, $d \eta = a^{-1} dt$, and in terms of the
re-scaled fields:
%%%%%%%%%%%%%%%%%
\beqn
\nonumber  \phi (t) &=& \frac{1}{\sqrt{\lambda} \> a(\eta)} \varphi (\eta)
\>\>,\>\>
\chi_{p \ell m} (t,\> r,\> \Omega) = \frac{1}{a (\eta)} X_p (\eta) Y_{p
\ell m} (r, \> \Omega) , \\
\Sigma (\eta) &=& \langle X^2 (\eta) \rangle .
\eeqn
The quantity $g^2 \Sigma (\eta)$ measures the growth of the back-reaction
upon $\varphi (\eta)$, and the spatial harmonics $Y_{p\ell m}$ obey:
%%%%%%%%%%%%%%%%%%
\beq
{\bf L}^2 Y_{p\ell m} = - \left( p^2 - K \right) Y_{p \ell m} .
\eeq
In this paper, we will only be concerned with the
continuous spectrum of modes in both the $K = 0$ and $K = -1$ cases, with
the eigenvalue $p^2$ in the range $0 \leq p^2 < \infty$.  With these
definitions, the
equations of motion may be rewritten:
%%%%%%%%%%%%%%%%%%%%%%%%%
\beqn
\nonumber  \left[ \frac{d^2}{d \eta^2} - \frac{1}{a} \frac{d^2 a}{d
\eta^2} + \varphi^2 + g^2 \Sigma \right] \varphi (\eta) &=& 0, \\
\left[ \frac{d^2}{d \eta^2} - \frac{1}{a} \frac{d^2 a}{d \eta^2} + p^2 - K
+ \frac{g^2}{\lambda} \varphi^2 \right] X_p (\eta) &=& 0 .
\label{eom}
\eeqn
It is convenient to define the frequencies $W_p (\eta)$ and $\omega_p
(\eta)$ as
%%%%%%%%%%%%%%%%%%
\beqn
\nonumber W_p^2 (\eta) &\equiv& p^2 - K - \frac{1}{a (\eta)} \frac{d^2
a(\eta)}{d \eta^2} , \\
\omega_p^2 (\eta) &\equiv& W_p^2 (\eta) + \frac{g^2}{\lambda} \varphi^2
(\eta) .
\label{freqs}
\eeqn
In this way, $W_p$ is the (time-dependent) \lq\lq natural frequency" for
the $X_p$ modes in the absence of interactions, and $\omega_p$ is the
(time-dependent) \lq\lq natural frequency" for these modes when subject to
the $g^2 \chi^2 \phi^2$ coupling. \\
\indent  One further simplification may be made for the case of preheating
with massless fields.  As demonstrated in \cite{DK97}, the scale factor
$a(\eta)$, averaged over a period of the inflaton field's oscillations,
behaves as if the universe were radiation-dominated (even though all the
fields are far from thermal equilibrium).  Thus, to a good approximation,
we have the relation
%%%%%%%%%%%%%%%%%%%
\beq
\frac{1}{a (\eta)}\frac{d^2 a (\eta)}{d \eta^2} = - K
\label{aprimeprime}
\eeq
for the entire preheating epoch.  (We thus ignore for now the possibility
of additional resonance effects arising from the oscillation of $a(t)$;
see \cite{bassett}.)  This means that the Ricci curvature scalar vanishes
during preheating: 
%%%%%%%%%%%%%%%%%%
\beq
R (\eta) = \frac{6}{a^2 (\eta)} \left[ \frac{1}{a} \frac{d^2 a}{d \eta^2} 
+ K \right] \rightarrow 0 .
\label{R}
\eeq
Models of preheating in an expanding universe with
massless fields are thus conformal to models of
preheating in Minkowski spacetime, even though the fields are
minimally-coupled to $R$ rather than conformally-coupled to $R$. \\
\indent  It remains to consider appropriate initial
conditions for the
modes $X_p (\eta)$; this is crucial to any study of preheating, since the
initial conditions for $X_p$ come from canonically quantizing the field
$\chi$ on a proper Cauchy surface and choosing an appropriate vacuum state
around which the quantized field may be expanded.  As discussed in
\cite{DK97}, when
treating only the continuum of modes with $p^2 \geq 0$, the proper  
initial conditions for $X_p (\eta)$ take the same form for both $K = 0$
and $K = -1$.  There are two independent concerns for choosing an
appropriate vacuum state:  the nonequilibrium interactions amongst the
fields
(relevant even for discussions of preheating in Minkowski spacetime), and
the usual ambiguity regarding physical vacua in general-relativistic
settings.  We choose to study preheating as measured against the \lq\lq
adiabatic" vacuum:  both the transition from de Sitter spacetime and the
\lq\lq turning on" of the interaction between the fields are assumed to
occur adiabatically.  In the absence of interactions, the transition from
de Sitter spacetime to the more general FRW spacetime at the beginning of
preheating would yield the following initial condition for the $X_p$ modes
at the onset of preheating (taken to be the time $\eta = \eta_0$):  $X_p
(\eta_0) = [2 W_p (\eta_0)]^{-1/2}$ and $(d X_p/d \eta)_{\eta = \eta_0} =
- i [ W_p (\eta_0) /2]^{1/2}$.  These modes would represent the
free-particle \lq\lq adiabatic" states \cite{BirDavies}.  If the
interactions were also turned on adiabatically beginning some time before
$\eta_0$, then these initial conditions would be replaced by:
%%%%%%%%%%%%%%%%%%
\beq
X_p (\eta_0) = \frac{1}{\sqrt{2 \omega_p (\eta_0)}} \>\>,\>\>
\left(\frac{d X_p}{d \eta} \right)_{\vert_{\eta = \eta_0}} = -i \sqrt{
\frac{\omega_p (\eta_0)}{2}} .
\label{initcond}
\eeq
These give the initial conditions for the \lq\lq adiabatic" particle
states for the nonequilibrium dynamics at the onset of preheating, at
$\eta = \eta_0$.  With this choice of vacuum state, the number of produced
quanta per mode may similarly be written in the same form for both $K =
0$ and $K = -1$ \cite{DK97}:
%%%%%%%%%%%%%%%%%%%%
\beq
N_p = \frac{\omega_p (\eta)}{2} \left[ \left| X_p \right|^2 +
\frac{1}{\omega_p^2} \left| \frac{d X_p}{d\eta} \right|^2 \right] -
\frac{1}{2} .
\label{Np}
\eeq
Only when the $\chi$ field is properly quantized will this yield the
number of \lq\lq adiabatic"-state quanta produced per mode relative to the
initial Fock space vacuum.  It is clear that when $p$ lies in a
resonance band and $X_p \sim e^{\mu_\kappa \eta}$, then $N_p \sim e^{2
\mu_\kappa \eta}$, giving rise to the very efficient transfer of energy
from the oscillating $\varphi$ field into $\chi$ quanta.  \\
\indent  Finally, there is the issue of renormalization.  With our
choice of vacuum (and hence of initial conditions for $X_p$), the term
$\langle X^2 (\eta_0) \rangle$ is formally quadratically divergent.  This
ultraviolet divergence may be treated in either of two ways:  we may 
undertake a formal renormalization, as done, for example, in \cite{DBfrw} 
(see also \cite{Baacke}),
or we may use the \lq\lq physical" criterion that the energy density
contained in the vacuum fluctuations at the beginning of preheating must
be less
than that contained in the classical inflaton field, $\phi$, as done in
\cite{DK97}.  In practice, this amounts to setting $g^2
\Sigma (\eta_0) \ll \varphi_0^2 (\eta_0)$.  Calculationally, the
divergence will not be a problem,
because preheating generically produces many quanta in low-$p$ resonance
bands, and we will only be \lq following' the evolution of these
exponentially-growing unstable modes, so that in practice we will not have
to evaluate
a full integral over $0 \leq p < \infty$.  We may now turn to the
evolution of the fields during preheating.

\section{Evolution of the Fields during Preheating}
\indent  Using equation (\ref{aprimeprime}), the equation of motion for
$\varphi$ in equation (\ref{eom}) may be solved for early times, when
$g^2 \Sigma (\eta)$ may be neglected.  Setting $\eta_0 = 0$, the solution
is
\cite{DBan,DK97,GKLS}:
%%%%%%%%%%%%%%
\beq
\varphi (\eta) = \varphi_0 \> {\rm cn} \left( \sqrt{ \varphi_0^2 +
K}\>\eta,\> \varphi_0/\sqrt{2 (\varphi_0^2 + K )} \right) ,
\label{varphi}
\eeq
where ${\rm cn} (u, \nu)$ is the Jacobian cosine function, and
this
solution is valid for $\varphi_0^2 \geq 2 \vert K \vert$, appropriate for
chaotic inflation initial conditions.  Note that for non-zero $K$, the
opposite limit, $\varphi_0 \ll \vert K \vert$, will not produce any
resonant preheating effects:  in this case, $\varphi (\eta)$ reduces to
the non-periodic form $\varphi (\eta) \simeq \varphi_0 \cosh (\vert K
\vert \eta )$. \\
\indent  The solution in equation (\ref{varphi}) also
describes the early-time evolution of a massive inflaton for models in
Minkowski
spacetime:  in this case,
$K < 0$ corresponds to preheating with an explicit symmetry-breaking
potential, while $K > 0$ corresponds to preheating with a massive inflaton
but without symmetry breaking.  If we scale the fields by the inflaton
mass, then these reduce to $K = \pm 1$.  \cite{DBan}  For the remainder of
this paper, however, we will restrict attention to preheating in an
expanding universe with massless fields, and hence $K$ will equal either
$0$ or $-1$ depending on the spatial curvature of the metric.  \\
\indent  At this point, we could follow the procedure of
\cite{DBan,DK97} and substitute this solution for $\varphi (\eta)$ into
the equation of motion for the modes $X_p (\eta)$.  After much
calculation, explicit forms for the modes may then be found in terms of  
special functions for certain values of the ratio of couplings
$g^2/\lambda$.  Instead, we will adopt the methods developed in
\cite{GKLS}
to produce exact (early-time) solutions for $X_p$ by means of algebraic
construction.
This approach may be applied whenever the couplings satisfy $g^2/\lambda =
n (n + 1)/2$, with $n$ a positive integer.  With this ratio of the
couplings, the equation of motion for $X_p$ in equation (\ref{eom}) takes 
the form of a Lam\'e equation of order $n$.  \\
\indent  The authors of \cite{GKLS} study the case of $K = 0$.  With the
help of the following definitions, we may expand their method to the case
of non-zero $K$:
%%%%%%%%%%%%
\beq
\gamma \equiv \sqrt{\varphi_0^2 + K} \>\>,\>\> u \equiv \gamma \eta
\>\>,\>\> z \equiv {\rm cn}^2 (u, \nu) ,
\label{gamma}
\eeq
where $\nu = \varphi_0 / \sqrt{2 (\varphi_0^2 + K)} =
\varphi_0/\sqrt{2 \gamma^2}$.  The Jacobian elliptic functions obey the
relations (see, {\it e.g.}, \cite{Bateman})
%%%%%%%%%%%%%%
\beq
{\rm sn}^2 (u,\nu) + {\rm cn}^2 (u,\nu) = 1 \>\>,\>\> {\rm dn}^2 (u,\nu) +
\nu^2 {\rm sn}^2 (u,\nu) = 1 \>\>,\>\>
\frac{d}{du} {\rm cn} (u,\nu) = - {\rm sn} (u,\nu) \> {\rm dn} (u,\nu) ,
\eeq
so that
%%%%%%%%%%%%%%%%% 
\beq
\frac{dz}{du} = - 2 \left[ (1 - \nu^2) z + (2 \nu^2 - 1)
z^2 - \nu^2 z^3 \right]^{1/2} .
\eeq
Given the bound $\varphi_0^2 \geq 2 \vert K \vert$, the modulus
$\nu$ obeys
%%%%%%%%%%%%%%
\beq
\frac{1}{2} \leq \nu^2 \leq 1 .
\eeq
Deviations of $\nu^2$ from $1/2$ indicate effects from the
non-zero spatial curvature; the limit $\nu^2 \rightarrow 1/2$ is
the same as the limit $\varphi_0^2 \gg \vert K \vert$, that is, the limit
when the 
initial amplitude of the oscillating inflaton is much greater than the
associated curvature scale. \\
\indent  The Jacobian cosine function ${\rm cn} (u, \nu)$ is
periodic with period $4 K (\nu)$ (as measured in \lq $u$' units
of time), where $K (\nu)$ is the complete elliptic integral of
the first kind.  This means that $z (u)$ ($\propto\varphi^2 (u)$) has
period $2 K (\nu) \equiv 2 \omega$, and hence that Floquet's
theorem applies to the equation of motion for $X_p$ in equation
(\ref{eom}).  In terms of $z$, the equation of motion for $X_p$ becomes:
%%%%%%%%%%%%%%%%%%%
\beq
4 f(z) X_p^{\prime\prime} + 2 f^\prime (z) X_p^\prime + \left( \kappa^2 + 
2 \nu^2 \frac{g^2}{\lambda} z \right) X_p = 0 ,
\label{eom2}
\eeq
where primes denote $d/dz$, and we have defined
%%%%%%%%%%%%%%%%%
\beq
\kappa^2 \equiv p^2 / \gamma^2  .
\eeq
Finally, we have defined the function
%%%%%%%%%%%%%%%%
\beq
f(z) \equiv (1 - \nu^2) z + (2\nu^2 - 1) z^2 - \nu^2 z^3 .
\label{f}
\eeq
With these definitions, we may study equation (\ref{eom2}) for $X_p$ for  
any value of the couplings $g^2/ \lambda = n (n + 1)/2$, following the
general approach of \cite{GKLS}. \\
\indent  Because the differential equation for $X_p$ is second-order,
there will exist
two linearly-independent solutions, which we can label $U_1$ and $U_2$
(suppressing the index $p$ for the moment).  Consider the product $M (z)
\equiv U_1 (z) U_2 (z)$.  After straightforward algebra, and making use of
equation (\ref{eom2}), one can show that this function obeys the  
third-order differential equation:
%%%%%%%%%%%%%%%%%
\beq
 2f(z)M^{\prime\prime\prime}(z) + 3 f^\prime (z)
M^{\prime\prime}(z) 
+ \left[ f^{\prime\prime}(z) + 2 \left( \kappa^2 + 2\nu^2 
\frac{g^2}{\lambda}
z \right)\right] M^\prime (z) + 2\nu^2 \frac{g^2}{\lambda} M(z) = 0 .
\label{M}
\eeq
When $g^2/\lambda = n (n + 1)/2$, this equation may be solved in terms of 
polynomials in $z$ of order $n$, which we will label $M_{(n)}(z)$:
%%%%%%%%%%%%%%%%%%%
\beq
M_{(n)} (z) = \sum_{i=0}^n a^{(n)}_i z^{(n-i)},
\label{M2}
\eeq
and we will set $a^{(n)}_0 = 1$ for all $n$.  Because $z (u)$ is periodic
in $u$ with period $2 \omega = 2 K (\nu)$, $M_{(n)} (z)$ will
also be periodic in $u$.  We
will give some explicit examples below for $n = 1$ and $n = 2$.  With
these solutions for $M_{(n)}(z)$ in hand, one can then find solutions for
$U_1 (z)$ and $U_2 (z)$, using the equation for $X_p$ and the Wronskian
relation between $U_1$ and $U_2$.  \\
\indent  In general, for a differential equation
of the form
%%%%%%%%%%%%%%%%%%%   
\beq
\frac{d}{dx} \left[ A(x) \frac{dy}{dx} \right] + B(x) y = 0 ,
\eeq
the Wronskian for the two linearly-independent solutions will be
proportional to $1/A(x)$.  In our case, we have
%%%%%%%%%%%%%%%%%
\beq
\frac{d}{dz} \left[\sqrt{f(z)} \>\frac{d X_p}{dz} \right] + B(z) X_p = 0,
\eeq
and hence
%%%%%%%%%%%%%%%%%%%%%
\beq
U^\prime_1 U_2 - U_1 U_2^\prime = \frac{C_{(n)}}{\sqrt{f(z)}} ,
\label{Wron}
\eeq
where the constant $C_{(n)}$ will be a function of $\kappa$ and $\nu$ and
will depend on the order $n$.  Using this Wronskian and the fact that
$U_1(z) U_2(z) = M(z)$, we may write solutions for the two mode functions:
%%%%%%%%%%%%%%%%%%%
\beq
U_{1,2} (z) = N\sqrt{\vert M_{(n)} (z) \vert} \exp \left( \pm
\frac{C_{(n)}}{2} \int \frac{dz}{\sqrt{f(z)} M_{(n)}(z)} \right).
\label{U1U2}
\eeq
We have inserted a normalization factor $N$, which is set to unity in
\cite{GKLS}.  Instead, we will set 
%%%%%%%%%%%%
\beq
N =  \vert M_{(n)} (z = 1) \vert^{-1/2} ,
\label{N}
\eeq
since $z = 1$ at $u = 0$.  With this normalization, when the solutions are
written as functions of $u$, they satisfy $U_{1,2} (u = 0) = 1$; and, as
demonstrated in Appendix A for the case $n = 1$, they match the exact
solutions found
previously, by very different means. \\
\indent  With $U_{1,2} (u = 0) = 1$, the normalized mode solutions
$U_{1,2}$ do not obey the proper initial conditions at $u = \eta = 0$,
equation (\ref{initcond}).  We must instead take linear combinations of
these solutions \cite{DBan,DK97}:
%%%%%%%%%%%
\beq
X_p (u) = \frac{1}{2 \sqrt{2 \omega_p (0)}} \left[ \left(1 + i
\frac{\omega_p (0)}{\gamma U_1^\prime (u = 0)} \right) U_1 (u) + \left( 1
- i \frac{ \omega_p (0)}{\gamma U_1^\prime (u = 0)} \right) U_2 (u)
\right] ,
\label{XpU1U2}
\eeq
where primes here denote $d/du$.  Unless such
linear combinations of $U_{1,2}$ are taken, the number operator $N_p$ in
equation (\ref{Np}) will {\it not} measure particle-production with
respect to the physically-relevant initial vacuum state.  Numerically,
however, if one makes the approximation of only following the
exponentially-growing unstable modes,
then $N_p$ will only shift by a constant as one shifts the normalization
of $U_{1,2}$ and the linear combination for $X_p$. \\
\indent  The solutions $U_{1,2}$ will be exponentially unstable whenever
the exponent in equation (\ref{U1U2}) has a
non-zero real part; we will use this to determine the bounds on the
resonance bands in terms of $\kappa$ and $\nu$.  
Furthermore, given the periodicity of $\varphi (u)$ and the
quasi-periodicity of $U_{1,2} (u)$, the authors of \cite{GKLS}
demonstrate that the characteristic exponent may be written as a definite 
integral:
%%%%%%%%%%
\beq
\mu_\kappa = - \frac{C_{(n)}}{2 \omega} \int_0^1 \frac{dz}{\sqrt{f(z)}
M_{(n)} (z)} .
\label{muk}
\eeq
We will present means for further evaluating $\mu_\kappa$ given this
expression in Section IV; in particular, we will provide a transparent
demonstration that the definite integral in equation (\ref{muk}) is always
purely real, so that all resonance structure comes from the behavior of
$C_{(n)}$.  We will demonstrate explicitly in Appendix B that this form
for $\mu_\kappa$ does indeed match the exact solutions found earlier in
\cite{DBan,DK97} for the two specific cases already studied.  We will
show in Appendix A that the solutions in equation (\ref{U1U2}) obey
Floquet's theorem, and can be written as $U_{1,2} (u) = P(\pm u) \exp
(\mp \mu_\kappa u)$, with $P(u + 2\omega) = P (u)$.  \\
\indent  By plugging the solutions for $U_{1,2}$ in equation (\ref{U1U2}) 
back into the equation of motion for $X_p$, one finds (dropping the
subscript $(n)$ for the moment):
%%%%%%%%%%%%%%%%%%%%%5
\beq
C^2 = M^{\prime 2} f - 2 M^{\prime\prime}M f - M^{\prime} M f^\prime -
\kappa^2 M^2 - \nu^2 n(n+1) M^2 z .
\eeq
It is straightforward to confirm that all of the $z$-dependence in this   
expression cancels exactly for any order $n$, leaving the simpler
relation:
%%%%%%%%%%%%%%%%%%%%%
\beq
C^2_{(n)} = -\kappa^2 \left(a_n^{(n)}\right)^2 - a_{n-1}^{(n)} a_n^{(n)}
(1 - \nu^2) .
\label{C}   
\eeq
Note that equations (\ref{eom2}), (\ref{f}), (\ref{M}), (\ref{Wron}), and 
(\ref{U1U2}) all reduce to the forms in \cite{GKLS} when $K = 0$ ($\nu^2
= 1/2$).  \\
\indent  For $g^2 = \lambda$ (that is, $n = 1$), equation (\ref{M}) may be
solved, yielding:
%%%%%%%%%%
\beqn
\nonumber  M_{(1)} (z) &=& z - \frac{\kappa^2}{\nu^2} + \frac{1}{\nu^2} -
2 , \\ 
\nu^4 C_{(1)}^2 &=& - \kappa^6 + 2 \kappa^4 (1 - 2 \nu^2) - \kappa^2 (1 -
5 \nu^2 + 5 \nu^4 ) - \nu^2 (1 - 3 \nu^2 + 2 \nu^4) .
\eeqn
When $K = 0$ ($\nu^2 = 1/2$), these become $M_{(1)} (z) \rightarrow z - 2
\kappa^2$ and $C_{(1)}^2 \rightarrow \kappa^2 (1 - 4 \kappa^4)$.  Thus, in
this limit, there will exist one single resonance band for positive
$\kappa^2$, determined by when $C_{(1)}^2 > 0$, and given by
%%%%%%%%%%%
\beq
0 \leq \kappa^2 \leq \frac{1}{2} .
\eeq
This matches the resonance structure found in both \cite{DBan,GKLS}.  It
is interesting that in the opposite limit, with non-zero $K$ and
$\varphi_0^2 \sim {\cal O} (\vert K \vert)$ ($\nu^2 \rightarrow 1$),
$C_{(1)}^2 \rightarrow - \kappa^6 - 2 \kappa^4 - \kappa^2$, and there will
not exist {\it any} resonance bands for $\kappa^2 \geq 0$. In fact, in an
open universe, there will exist a minimum value of $\varphi_0$ for any
given ratio $g^2 / \lambda$ below which no modes with $\kappa^2 \geq 0$
will be resonantly amplified.  This issue will be treated further in
\cite{DKinprep}. \\
\indent  For the next-simplest case, $g^2 = 3 \lambda$ ($n = 2$), it will
be easier to write the solutions in terms of the coefficients of the
expansion in equation (\ref{M2}) (recalling that $a_0^{(2)} = 1$):
%%%%%%%%%%%
\beqn
\nonumber M_{(2)} (z) &=& z^2 + a_{1}^{(2)} z + a_{2}^{(2)} , \\
\nonumber  a^{(2)}_{1} &=& -\frac{1}{3 \nu^2} \left[ \kappa^2 - 4 ( 1 - 2
\nu^2) \right] , \\
\nonumber  a^{(2)}_{2} &=& \frac{1}{9 \nu^4} \left[ \kappa^4 - 5 \kappa^2
(1 - 2 \nu^2) + 4 - 25 \nu^2 + 25 \nu^4 \right] , \\
C_{(2)}^2 &=& - \kappa^2 \left( a^{(2)}_2 \right)^2 - a^{(2)}_1 a^{(2)}_2
(1 - \nu^2) .
\label{Mn=2}
\eeqn
When $K = 0$, this reduces to 
%%%%%%%%%%%
\beq
C_{(2)}^2 \rightarrow \frac{16}{81} \kappa^2 \left( \kappa^4 -
\frac{9}{4} \right) \left(3 - \kappa^4 \right) ,
\eeq
revealing the presence of a single resonance band for positive $\kappa^2$:
%%%%%%%%%%%%
\beq
\frac{3}{2} \leq \kappa^2 \leq \sqrt{3} ,
\eeq
matching the resonance structure found in \cite{DK97,GKLS}.  Note that in
the opposite limit, $\nu^2 \rightarrow 1$, 
%%%%%%%%%
\beq
C_{(2)}^2 \rightarrow -\frac{1}{81} \kappa^2 \left(\kappa^4 + 4 \right)^2
,
\eeq
again revealing {\it no} resonance bands for $\kappa^2 \geq 0$. \\
\indent  We now turn to the treatment of the characteristic exponents,
$\mu_\kappa$.

\section{Evaluating the Characteristic Exponents}
\indent  As noted above, the authors of \cite{GKLS} found a form for the
characteristic exponent, $\mu_\kappa$:
%%%%%%%%%%
\beq
\mu_\kappa = - \frac{C_{(n)}}{2 \omega} \int_0^1 \frac{dz}{\sqrt{ f(z)}
M_{(n)} (z)} .
\eeq
In this section, we will rewrite the definite integral in a way
which demonstrates explicitly that the integral is real for all
$n$ and $\nu$, so that the resonance bands are determined entirely by
where the constant $C_{(n)}$ has non-zero real parts.  In addition,
the integral may be evaluated in closed form in many cases.  Such
closed-form solutions for $\mu_\kappa$ may then be evaluated without
numerical integration, and will facilitate comparison with known
solutions, as discussed in the appendices. \\
\indent  With the substitution
%%%%%%%%%%%%
\beq
z = 1 - \sin^2 \theta,
\eeq
the indefinite integral in equation (\ref{U1U2}) becomes:
%%%%%%%%%%%
\beq
-\int \frac{dz}{\sqrt{f (z)} M_{(n)} (z)} =  2 \int \frac{d
\theta}{\left[1
- \nu^2 \sin^2 \theta \right]^{1/2} M_{(n)} (\sin^2 \theta)} .
\label{theta3}
\eeq
Thus, given $1/2 \leq \nu^2 \leq 1$, both the integral over $dz$ and the
integral over $d\theta$ will be manifestly real for all $n$ and
$\nu$.  \\
\indent  The form of the integral over $d \theta$ in equation
(\ref{theta3}) further encourages comparison with the incomplete elliptic
integral of the third kind (see, {\it e.g.}, \cite{AbSteg}):
%%%%%%%%%%%%%
\beq
\Pi (n; \>\varphi \backslash \alpha) = \int^\varphi d\theta \left[ 1 - n
\sin^2 \theta \right]^{-1} \left[ 1 - \sin^2 \alpha \sin^2 \theta
\right]^{-1/2} .
\label{Pi}
\eeq
Whenever the function $M_{(n)} (\sin^2 \theta)$ may be factored into $n$
real roots, the integral over $dz$ may be written explicitly as a sum over
$n$ distinct $\Pi$-functions.  In the cases when $M_{(n)} (\sin^2 \theta)$
cannot be so factored, then the (principal value of the) integral over
$d\theta$ in equation
(\ref{theta3}) will still be well-defined over the limits of integration
$(z: 0, \> 1) \rightarrow (\theta: \pi/2 , \> 0)$. \\
\indent  Consider the case when $M_{(n)} (\sin^2 \theta)$ can be factored
into $n$ real roots; writing $x \equiv \sin^2 \theta$, we have:
%%%%%%%%%%%%
\beq
\frac{1}{M_{(n)} (x)} = \frac{1}{\prod_{i = 1}^n \left( \beta_i - x
\right)} = \sum_{i = 1}^n \frac{D_i}{\left(1 - \beta_i^{-1} x \right)} ,
\label{product}
\eeq
where the $n$ constant coefficients $D_i$ may be determined by the set of
$n$ equations:
%%%%%%%%%%%%
\beqn
\nonumber  \sum_{i = 1}^n D_i \left( \prod_{j = 1}^n \beta_j \right) &=& 1
, \\
\nonumber D_1 \left( \beta_2^{-1} + \beta_3^{-1} + ... + \beta_n^{-1}
\right) + {\rm c.p.} &=& 0 , \\
\nonumber  D_1 \left( \beta_2^{-1} \beta_{3}^{-1} + \beta_2^{-1}
\beta_4^{-1} + ... + \beta_{n-1}^{-1} \beta_{n}^{-1} \right) + {\rm c.p.}
&=& 0 , \\
\nonumber  ... \\
D_1 \left( \beta_2^{-1} \beta_3^{-1} \times ... \times \beta_n^{-1}
\right) + {\rm c.p.} &=& 0 ,
\label{Di}
\eeqn
where ${\rm c.p.}$ denotes all cyclic permutations.  Note that when the
constant portion of $M_{(n)} (\sin^2 \theta)$ (that is, $a_{n}^{(n)}$) is
non-zero, all of the roots $\beta_i$ will be non-zero, and hence their
inverses $\beta_i^{-1}$ will always be well defined.  In general, both the
constants $D_i$ and $\beta_i$ will be functions of $n$, $\kappa^2$, and
$\nu$. We
will give explicit examples of the factorization of equation
(\ref{product}) for $n = 1$ and $n = 2$ below. \\
\indent  In the cases when $M_{(n)} (\sin^2 \theta)$ admits such a
factorization, then we may write the indefinite integral in the exponent
of $U_{1,2}$ as:
%%%%%%%%%%%%%%%%
\beq
-\int \frac{dz}{\sqrt{f (z)} M_{(n)} (z)} =  2 \sum_{i = 1}^n D_i \Pi
\left(\beta_i^{-1};\> \arcsin \sqrt{1-z} \>
\backslash \arcsin \nu \right) .
\label{int}
\eeq
Using the fact that $\Pi (n;\> 0 \> \backslash \alpha ) = 0$ and $\Pi (n;
\>
\pi/2 \>\backslash \alpha) = \Pi (n \backslash \alpha)$ (the complete
elliptic integral of the third kind), we may further evaluate the
characteristic exponent, $\mu_\kappa$:
%%%%%%%%%%%%%%
\beq
\mu_\kappa = - \frac{C_{(n)}}{2 \omega} \int_0^1 \frac{dz}{
\sqrt{f (z)} M_{(n)} (z)} 
= - \frac{C_{(n)}}{\omega} \sum_{i = 1}^n D_i \Pi \left(\beta_i^{-1}
\backslash \arcsin \nu \right) .
\label{muk2}
\eeq
For cases in which a given $\beta_j^{-1} > 1$, one may always rewrite $\Pi
(\beta_j^{-1} \backslash \arcsin \nu)$ in terms of a sum of
complete elliptic integrals, each of which is purely real (see
\cite{AbSteg}). 
This closed-form solution for $\mu_\kappa$, valid whenever $M_{(n)}
(\sin^2 \theta)$ has the $n$ non-zero, real roots $\beta_i$, may thus be
evaluated in terms of well-known functions, without any need for numerical
integration. \\
\indent   Consider the simplest case, $n = 1$.  In this case, $M_{(1)}
(\sin^2 \theta)$ may always be factored as needed:
%%%%%%%%%%%
\beq
M_{(1)} (\sin^2 \theta) =  - \sin^2 \theta - \frac{\kappa^2}{\nu^2} +
\frac{1}{\nu^2} (1 -  \nu^2) , 
\eeq
or,
%%%%%%%%%%%%
\beq
\beta_1^{-1} = D_1 = \frac{\nu^2}{1 - \nu^2 - \kappa^2}  .
\label{beta-D1}
\eeq
This yields
%%%%%%%%%%%%
\beq
\mu_\kappa = - \frac{C_{(1)}}{\omega} \left( \frac{\nu^2}{1 - \nu^2 -
\kappa^2} \right) \Pi
\left( \frac{ \nu^2}{1 - \nu^2 - \kappa^2} \> \backslash \> \arcsin
\nu \right) .
\label{mukn=1} 
\eeq
Equation (\ref{mukn=1}) reveals that preheating in an open universe can be
more efficient than in a spatially-flat one.  The maximum for $\mu_\kappa$
when $K = 0$ ($\nu^2 = 1/2$) is $\mu_\kappa = 0.147$ at $\kappa^2 = 0.228$
(matching the result found in \cite{GKLS}).  Yet $\mu_\kappa$ grows in the
range $1/2 < \nu^2 < 0.76$, reaching a maximum of $\mu_\kappa = 0.224$ at
$\kappa^2 = 0$ and $\nu^2 = 0.76$.  This qualitative difference in the
spectra will be treated further in \cite{DKinprep}.  We will demonstrate
below in Appendix B that the analytic solution for $\mu_\kappa$ in
equation (\ref{mukn=1}) matches
the exact solution found previously for $n = 1$, with $\nu^2 = 1/2$. \\
\indent  For $n = 2$, we have (again using $x \equiv \sin^2 \theta$):
%%%%%%%%%%
\beq
M_{(2)} (x) = x^2 - x \left( 2 + a_{1}^{(2)} \right) + \left( 1 +
a_1^{(2)} + a_2^{(2)} \right) , 
\eeq
or (dropping the superscript $(2)$):
%%%%%%%%%%%%%
\beqn
\nonumber  \beta_{1,2} &=& \frac{1}{2} \left[ 2 + a_1 \mp \sqrt{(a_1)^2 -
4 a_2} \right] \\
D_{1,2} &=& \frac{1}{2} \frac{1}{\left(1 + a_1 + a_2 \right) \sqrt{(a_1)^2
- 4 a_2} }\left[ \pm 2 \pm a_1 + \sqrt{(a_1)^2 - 4 a_2} \right] .
\label{mukn=2}
\eeqn
Note that when $(a_1)^2 < 4 a_2$, the definite integral in $\mu_\kappa$
will still be purely real; it simply will not be expressible as a
sum of complete elliptic integrals.   In the limit $\nu^2 \rightarrow
1/2$,
however, the condition $(a_1)^2 \geq 4a_2$ is equivalent to $\kappa^4 \leq
3$ (see equation (\ref{Mn=2})).  And in this limit, the single resonance
band extends only to
$\kappa^4 \leq 3$, so for the entire resonance band, the roots and
coefficients in equation (\ref{mukn=2}) will all be purely real. \\
\indent  As with the $n = 1$ case, preheating in an open universe with $n
= 2$ can be more efficient than in a flat one.  In this case,
the enhancement of the resonance can be far more dramatic:
whereas $\mu_\kappa$
reaches a maximum value when $K = 0$ ($\nu^2 = 1/2$) of $\mu_\kappa =
0.036$ (at $\kappa^2 = 1.615$), $\mu_\kappa$ rises sharply over the range
$1/2 < \nu^2 \leq 0.90$.  At $\nu^2 = 0.72$, $\mu_\kappa$ reaches a
maximum of $0.097$ (at $\kappa^2 = 0.64$), while at $\nu^2 = 0.90$,
$\mu_\kappa$ reaches a maximum of $0.193$ (at $\kappa^2 = 0$).  This
yields a characteristic exponent over five times greater than the maximum
reached in a spatially-flat universe.  We will
demonstrate analytically in Appendix B that for $n = 2$ and in the limit
$\nu^2 \rightarrow 1/2$, the closed-form solution obtained from equations
(\ref{muk2}) and (\ref{mukn=2}) again matches the exact solution. 

\section{Conclusions}
\indent  The methods developed in \cite{GKLS} for the analytic study of
the resonance structure of preheating with massless fields are
powerful and highly efficient.  We have extended their work here to cover
both models with non-zero spatial curvature (relevant to studies of open
inflation), and models in Minkowski spacetime with or without
explicit symmetry-breaking.  By considering a physically well-motivated
choice
of initial vacuum state, we have also fixed the overall normalization of
the mode functions, which provides a simple form for the number of
$\chi$ quanta produced per mode during preheating.  With this choice of
normalization, the solutions for the mode functions found here by simple
algebraic construction match the solutions found previously in terms of
special functions (see Appendix A). \\
\indent  Beyond the relative ease with which solutions for the mode
functions may be found using the methods of \cite{GKLS}, there is an added
benefit of this approach over that taken in, {\it e.g.}, \cite{DBan,DK97}:
the
characteristic exponent, $\mu_\kappa$, which determines the rate at which
particles are resonantly produced during preheating (with $\ln N_p \simeq 
2 \mu_\kappa u$), may be evaluated
independently of finding the full solutions for the mode functions,
$U_{1,2} (z)$.  We have shown here that in many cases of interest, these
$\mu_\kappa$ parameters may in fact be solved for exactly in closed form,
in terms of well-known functions.  Written in this form, the
characteristic exponents, calculated directly from equation (\ref{muk2}),
match the solutions found previously (see Appendix B). \\
\indent  Of course, this method, though simple and efficient, is still
limited to the cases in which the ratio of the couplings satisfies $g^2 /
\lambda = n (n + 1)/2$, with $n$ a positive integer.  The authors of
\cite{GKLS} show, by means of
numerical integration for arbitrary positive $g^2 / \lambda$, that
$\mu_\kappa$ is not a monotonic function of the couplings; for slightly
different values of $g^2 / \lambda$, the resonance can be
much stronger than it is with some of the integer-valued ratios studied
here.  Still, it is useful in general to have analytic solutions in hand,
especially if one wants to compare different models, such as $K = 0$
versus $K = -1$, or, for Minkowski spacetime, $m_\phi^2 > 0$ versus
$m_\phi^2 < 0$.  Such analytic comparisons have revealed here, for
example, that when $K = -1$ and the initial amplitude of the inflaton's
oscillations falls below a minimum value, {\it no} modes with $\kappa^2
\geq 0$ will be resonantly amplified, even though there is no associated
threshold in the $K = 0$ case.  Moreover, when $K = -1$, certain
regions of parameter space yield rates of resonant particle production 
over five times greater than the corresponding rates in a $K = 0$
universe.  These topics will be treated further in \cite{DKinprep}.

\section*{Acknowledgments}
\indent  This work was supported in part by the National Science
Foundation, grant PHY-92-18167.

\section*{Appendix A}
\indent  In this appendix, we demonstrate that the solutions $U_{1,2} [z
(u)]$ in equation (\ref{U1U2}), with the normalization of equation
(\ref{N}), match the exact solutions found in terms of special functions.
We will examine the simplest case, $n = 1$, which was studied in
\cite{DBan}.  In terms of the parametrization of equation (\ref{gamma}),
and setting $\nu^2 = 1/2$, we may follow the same steps as
in \cite{DBan} to arrive at the solution:
%%%%%%%%%%%%%%%%%%%%%%%%
\beq
U_\kappa (u) = \frac{\vartheta_1 \left(\frac{u}{2\omega} + v
\right)}{\vartheta_1 (v)} \frac{\vartheta_4 (0)}{\vartheta_4 \left(
\frac{u}{2\omega} \right)} \exp \left[ - u Z (2 \omega v) \right] ,
\label{Un1}
\eeq
and $U_{1,2} (u) = U_\kappa (\pm u)$.  Here $\vartheta_i (x)$ are the
Jacobian theta functions, $\omega = K (\nu)$ is the complete
elliptic integral of the first kind, and $Z (x)$ is the Jacobian zeta
function (see, {\it e.g.}, \cite{Bateman,AbSteg}).  The quantity $v$ is
related to $\kappa^2$ by \cite{DBan}:
%%%%%%%%%%%%%%%%%%%
\beq
\wp (2 \omega v + \omega^\prime) = - \kappa^2 ,
\label{v1}
\eeq
where $\wp (x)$ is the doubly-periodic Weierstrass function.  We have also
used 
$\omega^\prime \equiv i K^\prime (\nu)$, where $K^\prime
(\nu) = K (\nu^\prime)$, the complete elliptic integral
for the complementary modulus, $\nu^\prime \equiv (1 -
\nu^2 )^{1/2}$. 
\cite{Bateman,AbSteg}  We may rewrite the $v$-dependence in
terms of Jacobian elliptic functions, using
%%%%%%%%%%%%%%%%%
\beq
{\rm sn}^2 (u, \nu) = \frac{1}{\nu^2 {\rm sn}^2 (u +
\omega^\prime, \nu)} = 2 \wp (u + \omega^\prime ) + 1 ,
\label{sn-wp}
\eeq
where the last expression holds for $\nu^2 = 1/2$.  Making these
substitutions, equation (\ref{v1}) may be rewritten:
%%%%%%%%%%%%%%%%%%%%%%%%
\beq
1 - 2 \kappa^2 = {\rm sn}^2 (2 \omega v, \nu) .
\label{v2}
\eeq
Inside the resonance band, with $0 \leq \kappa^2 \leq 1/2$, we thus have
$0 \leq v \leq 1/2$. \\
\indent  In order to demonstrate the equivalence between the solutions in
equations (\ref{U1U2}) and (\ref{Un1}), we will first examine the term
outside of the exponent in equation (\ref{U1U2}):
%%%%%%%%%%%%%%%%%%%%%
\beq
\sqrt{ \vert M_{(1)} (z) \vert} = \sqrt{ z - 2 \kappa^2} = \left[ {\rm
sn}^2 (2 \omega v, \nu) - {\rm sn}^2 (u, \nu)
\right]^{1/2} .
\label{Msn}
\eeq
Again using equation (\ref{sn-wp}), the following relations between
the theta functions \cite{Bateman}
%%%%%%%%%%%%%%%%%%%%%%
\beqn
\nonumber \vartheta_1 \left( x + \frac{1}{2} \tau \right) &=& i \exp
\left[ - i \pi \left( x + \frac{1}{4} \tau \right) \right] \vartheta_4 (x)
, \\
\vartheta_4 \left( x + \frac{1}{2} \tau \right) &=& i \exp \left[ -i \pi
\left( x + \frac{1}{4} \tau \right) \right] \vartheta_1 (x) 
\label{theta1,4}
\eeqn
(where $\tau \equiv \omega^\prime / \omega$), and the relation (for
$\nu^2 = 1/2$)
%%%%%%%%%%%%%%%%%%%%%%
\beq
\wp (u) = -\frac{1}{2} + \frac{1}{4 \omega^2} \left[ \frac{
\vartheta_1^\prime (0)}{\vartheta_4 (0)} \frac{ \vartheta_4 \left(
\frac{u}{2 \omega} \right)}{\vartheta_1 \left( \frac{u}{2 \omega} \right)}
\right]^2 ,
\label{wp-theta}
\eeq
we may rewrite equation (\ref{Msn}) in terms of theta functions:
%%%%%%%%%%%%%%%%%%
\beq
\sqrt{ \vert M_{(1)} (z) \vert} = \frac{1}{\sqrt{2} \> \omega}
\frac{\vartheta_1^\prime
(0)}{\vartheta_4 (0)} \frac{1}{\vartheta_4 (v) \vartheta_4
(\frac{u}{2\omega}) }
\left[ \vartheta_1^2 (v) \vartheta_4^2 (\frac{u}{2\omega}) -
\vartheta_1^2 (\frac{u}{2\omega}) \vartheta_4^2 (v) \right]^{1/2} .
\eeq
The normalization of equation (\ref{N}) may be similarly re-expressed:
%%%%%%%%%%%%%%%%%%%%%
\beq
N = \frac{1}{\sqrt{1 - 2 \kappa^2}} = \frac{1}{{\rm sn} ( 2 \omega v,
\nu )} .
\eeq
Finally, using the relation between the squares of the theta functions
\cite{Whit}:
%%%%%%%%%%%%%%%%%%%%%%
\beq
\vartheta_1^2 (y) \vartheta_4^2 (z) - \vartheta_1^2 (z) \vartheta_4^2 (y)
 = \vartheta_1 (y + z) \vartheta_1 (y - z) \vartheta_4^2 (0) ,
\eeq
we find
%%%%%%%%%%%%%%%%%%%%%%%%
\beq
N \sqrt{ \vert M_{(1)} (z) \vert} = \frac{ \vartheta_4 (0)}{\vartheta_1
(v)
\vartheta_4 (\frac{u}{ 2\omega})} \left[ \vartheta_1 \left(v + \frac{u}{2
\omega}
\right) \vartheta_1 \left( v - \frac{u}{2 \omega} \right) \right]^{1/2} .
\label{NM1}
\eeq
Now we may turn to the exponent in equation (\ref{U1U2}). \\
\indent  Using equations (\ref{int}), (\ref{beta-D1}), (\ref{sn-wp}), and
(\ref{v2}),
with
$\nu^2 = 1/2$, we may write the exponent of equation (\ref{U1U2})
as
%%%%%%%%%%%%%%%%%%%%%%%%
\beq
\frac{C_{(1)}}{2} \int \frac{dz}{\sqrt{f (z)} M_{(1)} (z)} = - C_{(1)} D_1
\Pi \left( \frac{1}{2} {\rm sn}^2 (2 \omega v + \omega^\prime); \> \arcsin
\sqrt{1 - z} \> \backslash \> \arcsin (\nu) \right) .
\label{exponent1}
\eeq
Because $0 < \frac{1}{2} {\rm sn}^2 (2 \omega v + \omega^\prime) \leq
\nu^2$, we may rewrite the incomplete elliptic integral as
follows \cite{AbSteg}:
%%%%%%%%%%%%%%%%%%%%%%%%%%
\beq
\Pi (n; \> \varphi \> \backslash \alpha ) = \delta_1 \left[ - \frac{1}{2}
\ln \left(\frac{ \vartheta_4 (y + \beta)}{\vartheta_4 (y - \beta)} \right)
+ y \frac{ \vartheta_1^\prime (\beta)}{\vartheta_1 (\beta)} \right] ,
\label{Pi-theta}
\eeq
with
%%%%%%%%%%%%%%%%%%%%%%%
\beqn
\nonumber  \epsilon &=& \arcsin ( n / \sin^2 \alpha)^{1/2} \>\>,\>\> \beta
= F (\epsilon \backslash  \alpha ) / 2 \omega , \\
y &=& F (\varphi  \backslash  \alpha) / 2 \omega \>\>,\>\> \delta_1 =
\left[ n (1 - n)^{-1} ( \sin^2 \alpha - n)^{-1} \right]^{1/2} ,
\label{epsilon}
\eeqn
where $F (\varphi \backslash \alpha)$ is the incomplete elliptic integral
of the first kind.
Here we have used the \lq $\pi$' conventions for the arguments of the
theta
functions of \cite{Bateman}.  After some straightforward algebra, equation
(\ref{exponent1}) may then be rewritten:
%%%%%%%%%%%%%%%%%%%%%
\beq
\frac{C_{(1)}}{2} \int \frac{dz}{\sqrt{f(z)} M_{(1)}(z)} = \frac{1}{2} \ln
\left( \frac{ \vartheta_1 (v + \frac{u}{2\omega})}{\vartheta_1 (v -
\frac{u}{2\omega})}\right) - \frac{u}{2\omega} \frac{\vartheta_4^\prime
(v)}{\vartheta_4 (v)} ,
\label{exponent2}
\eeq
after use has been made of equation (\ref{theta1,4}) and the fact that
$\vartheta_4 (-x) = \vartheta_4 (x)$.  Finally, using \cite{Bateman}
%%%%%%%%%%%%%%%%%%%%%
\beq
\frac{\vartheta_4^\prime (x)}{\vartheta_4 (x)} = 2 \omega Z (2 \omega x) ,
\eeq
and combining equations (\ref{NM1}) and (\ref{exponent2}), we find that
the solution for $U_1 [z(u)]$ found in Section III, equation (\ref{U1U2}),
may be rewritten as
%%%%%%%%%%%%%%%%%%%%%%%%%
\beq
U_1 (u) = \frac{\vartheta_1 (\frac{u}{2\omega} + v)}{\vartheta_1 (v)}
\frac{\vartheta_4 (0)}{\vartheta_4 (\frac{u}{2\omega})} \exp \left[ - u Z
(2 \omega v ) \right] .
\eeq
Given the periodicity of the theta functions, this is now in the form:
$U_1 (u) = P(u) \exp \left[ - \mu_\kappa u \right]$, with $P (u + 2
\omega) = P(u)$.  The second independent solution can then be written $U_2
(u) = P (-u) \exp \left[ \mu_\kappa u \right]$, and represents the growing
solution. \\
\indent  This completes our demonstration that the solutions (for $n = 1$)
found
here by means of very
simple algebraic construction, following the method of \cite{GKLS}, match
the exact solutions found earlier by much more difficult means
\cite{DBan}.  The comparison for $n = 2$ between the
solutions found in Section III and those found earlier in \cite{DK97}
follows very similarly.

\section*{Appendix B}
\indent  Whereas in Appendix A we demonstrated that the solutions $U_{1,2}
(z)$ for the mode functions match the earlier known solutions, in this
appendix we will demonstrate that the crucial quantities, $\mu_\kappa$,
as calculated directly from equation (\ref{muk2}) also agree with earlier
known solutions. The
difference between these two demonstrations lies in the fact that the
solutions $U_{1,2} (z)$ are defined in terms of an {\it indefinite}
integral over $z$, as in equation (\ref{U1U2}); the characteristic
exponents $\mu_\kappa$, however, are defined as {\it definite} integrals.
As noted in the conclusions, one benefit (among many) of following the
methods of \cite{GKLS} is that the $\mu_\kappa$ parameters may be
evaluated independently of the mode functions $U_{1,2} (z)$, which is not
possible if proceeding as in \cite{DBan,DK97}. \\
\indent  Consider the case $n = 1$ with $\nu^2 \rightarrow 1/2$.  Equation
(\ref{mukn=1}) then becomes:
%%%%%%%%%%%%%%%%%%%%%%%
\beq
\mu_\kappa = - \frac{1}{\omega} \sqrt{ \kappa^2 \left(\frac{  1 + 2
\kappa^2}{1 - 2 \kappa^2} \right)} \Pi \left( \frac{1}{1 - 2 \kappa^2}
\backslash \arcsin (\nu) \right) .
\label{mukB1}
\eeq
Written in this form, $\beta_1^{-1} > 1$, so we may rewrite the elliptic
integral as follows \cite{AbSteg}:
%%%%%%%%%%%%%%%%%%%%%%%
\beq
\Pi ( \beta_1^{-1} \backslash \alpha ) = \omega - \Pi (N\backslash \alpha)
= - \delta_1 \omega Z (\epsilon \backslash \alpha) ,
\label{PiN}
\eeq
with $N = \beta_1 \sin^2 \alpha$, and $\delta_1$ and $\epsilon$ defined as
in equation (\ref{epsilon}), now in terms of $N$ instead of $n$.  Then,
using equation (\ref{v2}), we have 
%%%%%%%%%%%%%%%%%%%
\beq
\epsilon = \arcsin \sqrt{1 - 2 \kappa^2} = \arcsin \left( {\rm sn} (2
\omega v ) \right).
\eeq
Noting that when $\varphi = \arcsin \left( {\rm sn} (x, \nu)
\right)$ \cite{AbSteg},
%%%%%%%%%%%%%%%%%
\beq
Z (\varphi \backslash \arcsin (\nu)) = Z (x \vert \nu^2)
= Z (x) ,
\eeq
we find that
%%%%%%%%%%%%%%%%%%%%%%%
\beq
\mu_\kappa = Z ( 2 \omega v)
\eeq
for $n = 1$ and $\nu^2 \rightarrow 1/2$.  This matches the result found in
\cite{DBan}. \\
\indent  For $n = 2$ and $\nu^2 \rightarrow 1/2$, we may compare with the
solution found in \cite{DK97}:
%%%%%%%%%%%%%%%%%%%%%%%
\beq
\mu_\kappa = \frac{1}{2 \omega} \left( \frac{ \vartheta_1^\prime
(\frac{a}{2\omega})}{\vartheta_1 (\frac{a}{2 \omega})} + \frac{
\vartheta_1^\prime ( \frac{b}{2 \omega})}{\vartheta_1 (\frac{b}{2\omega})}
\right) ,
\label{mukB2}
\eeq
where the constants $a$ and $b$ are defined by the relations
%%%%%%%%%%%%%%%%%%%%%%%
\beqn
\nonumber \wp (a) &=& - \frac{1}{6} \kappa^2 - \frac{1}{2} \sqrt{ 1 -
\frac{1}{3} \kappa^4} , \\
\wp (b) &=& - \frac{1}{6} \kappa^2 + \frac{1}{2} \sqrt{1 - \frac{1}{3}
\kappa^4} .
\eeqn
Writing these in terms of Jacobian elliptic functions, using the
identities employed in Appendix A, we have
%%%%%%%%%%%%%%%%%%%%%%%
\beqn
\nonumber {\rm sn}^2 (a - \omega^\prime) &=& 1 - \frac{1}{3} \kappa^2 -
\sqrt{1 -\frac{1}{3} \kappa^4} , \\
{\rm sn}^2 (b - \omega^\prime) &=& 1 - \frac{1}{3} \kappa^2 + \sqrt{1 -
\frac{1}{3} \kappa^4} .
\eeqn
Examination of equations (\ref{Mn=2}) and (\ref{mukn=2}) in the limit
$\nu^2 \rightarrow 1/2$ reveals that
%%%%%%%%%%%%%%%%%%%%%%
\beqn
\nonumber  \beta_1^{-1} &=& \frac{1}{ {\rm sn}^2 (a - \omega^\prime)} =
\frac{1}{2} {\rm sn}^2 (a) , \\
\beta_2^{-1} &=& \frac{1}{ {\rm sn}^2 (b - \omega^\prime)} = \frac{1}{2}
{\rm sn}^2 (b) .
\eeqn
Then
%%%%%%%%%%%%%%%%%%%%%%
\beqn
\nonumber \mu_\kappa &=& \sqrt{\frac{3}{4 \kappa^2} \left( \frac{\kappa^2
+ \frac{3}{2}}{\kappa^2 - \frac{3}{2}} \right)} \\
&\times& \left[ \left(-1 + \frac{1}{3} \kappa^2 - \sqrt{1 - \frac{1}{3}
\kappa^4} \right) \Pi \left(\frac{1}{2} {\rm sn}^2 (a) \backslash \arcsin
(\nu) \right) + \left(1 - \frac{1}{3} \kappa^2 - \sqrt{1 -
\frac{1}{3} \kappa^4} \right) \Pi \left( \frac{1}{2} {\rm sn}^2 (b)
\backslash \arcsin (\nu) \right) \right] .
\eeqn
Since these $\beta_i^{-1}$ each satisfy $0 \leq \beta_i^{-1} \leq
1/2$, we may again use equation (\ref{Pi-theta}) for each of the two
complete elliptic integrals.  In each case, the $\ln ( \vartheta_4 (y +
\beta)/\vartheta_4 (y - \beta))$-term vanishes, leaving, after much
algebra,
%%%%%%%%%%%%%%%%%%%%%%%%%%
\beq
\mu_\kappa = \frac{1}{2 \omega} \left( \frac{ \vartheta_1^\prime
(\frac{a}{2\omega})}{\vartheta_1 (\frac{a}{2\omega})} + \frac{
\vartheta_1^\prime (\frac{b}{2\omega})}{\vartheta_1 (\frac{b}{2\omega})}
\right) .
\eeq
Thus, the closed-form solution for $\mu_\kappa$ found in Section IV,
equation (\ref{muk2}), matches the solution found previously in
\cite{DK97} for $n = 2$.

%_________________________________
  
\end{document}